\documentclass[conference]{IEEEtran}
\IEEEoverridecommandlockouts

\usepackage{amsmath,amssymb,amsfonts}
\usepackage{bm}
\usepackage{algorithmic}
\usepackage{algorithm}
\usepackage{graphicx}
\usepackage{textcomp}
\usepackage{xcolor}
\usepackage{booktabs}
\usepackage{cite}
\usepackage{multirow}

\def\BibTeX{{\rm B\kern-.05em{\sc i\kern-.025em b}\kern-.08em
    T\kern-.1667em\lower.7ex\hbox{E}\kern-.125emX}}

\begin{document}

\title{Leakage-Aware Encrypted Semantic Communication with Key-Conditioned Signal Perturbation
  \thanks{This work is supported by the American University of Beirut
    University Research Board (URB) and Vertically Integrated Projects (VIP) Program.}
}
\author{
\IEEEauthorblockN{Omar Kaaki, Maria Slim, Mariette Awad, Hadi Sarieddeen}
\IEEEauthorblockA{Department of Electrical and Computer Engineering \\
American University of Beirut, Beirut, Lebanon \\
Email: \{ohk06, mas194\}@mail.aub.edu, \{mariette.awad, hadi.sarieddeen\}@aub.edu.lb}
}

\maketitle

\begin{abstract}
Encrypted semantic communication systems protect transmitted messages by inserting key-conditioned modules into a deep-learning semantic communication pipeline, but existing designs rely on a fixed adversarial weight and operate mainly at the semantic-feature level. We propose two changes to the encrypted semantic communication system (ESCS): a key-conditioned signal-domain perturbation layer, called learned adversarial perturbation injection (LAPI), and an information noise-contrastive estimation (InfoNCE)-guided adaptive training rule that {adds an adaptive leakage penalty alongside the fixed adversarial objective} using an InfoNCE-based proxy for mutual information leakage. Experiments on the Europarl corpus over additive white Gaussian noise (AWGN) and Rayleigh fading channels show that InfoNCE-guided training is the primary source of eavesdropper suppression, reducing the eavesdropper's bilingual evaluation understudy (BLEU) score to approximately $0.07$ at high signal-to-noise ratio (SNR). Adding LAPI to the InfoNCE-guided objective further lowers the InfoNCE-estimated leakage by $5$--$9\%$ with {low measured inference overhead}, while its impact on legitimate-receiver BLEU depends on channel conditions. The two mechanisms operate on different parts of the pipeline and can be jointly applied.
\end{abstract}

\section{Introduction}

Semantic communication transmits message meaning rather than exact bit sequences, achieving strong reconstruction at low signal-to-noise ratios (SNRs) through joint source-channel coding (JSCC)~\cite{deepsc,qin2022}. Deep-learning systems such as DeepSC~\cite{deepsc} use Transformer architectures to learn compact semantic representations robust to channel noise.

Because wireless transmissions are exposed to unintended receivers, semantic latents must be treated as potentially observable rather than private. Luo~\emph{et~al.}~\cite{escs} addressed this with the encrypted semantic communication system (ESCS), which adds symmetric encryption and decryption modules to DeepSC and trains them adversarially against a simulated eavesdropper. ESCS reduces eavesdropper reconstruction quality, but its key-conditioned protection operates at the semantic-feature level, while the weight balancing legitimate-receiver utility and eavesdropper degradation remains fixed during training. Two limitations motivate our work. \emph{(L1) Single-domain protection:} the transmitted signal lacks an additional key-conditioned signal-domain protection mechanism. \emph{(L2) Fixed and indirect privacy objective:} the utility--privacy weight is fixed, and the privacy signal relies on eavesdropper cross-entropy rather than {a direct training-time estimate of statistical dependence between the unencrypted semantic representation and the eavesdropper's representation}.


ESCS~\cite{escs} inserts lightweight multilayer perceptron (MLP) encryptor/decryptor modules into DeepSC~\cite{deepsc} and trains them adversarially against an eavesdropper. Earlier work showed that neural networks can protect communications through adversarial neural cryptography~\cite{abadi2016}. Recent studies explore complementary privacy and security mechanisms. A cooperative adversarial perturbation layer adds a learned perturbation to the transmitted waveform to reduce semantic leakage~\cite{sagduyu2025}. Learnable differential-privacy noise for image semantic communication was studied in~\cite{shao2025}, while diffusion-based modules add and remove artificial noise~\cite{he2024}. Invertible-neural-network steganography embeds private signals into non-sensitive host transmissions~\cite{tang2025}. SignDeepSC~\cite{alhaj2026signdeepsc} addresses physical-layer man-in-the-middle attacks using a compact semantic signature over an auxiliary channel and a signature-assisted decoder. Our work extends ESCS with key-conditioned signal-domain perturbation and leakage-aware training, {operating at transmitted-signal and training-objective levels}.

Mutual information neural estimation (MINE)~\cite{mine} provides a differentiable lower bound on mutual information (MI) based on the Donsker--Varadhan representation, but its estimation variance can grow substantially as the underlying MI increases~\cite{smile}. InfoNCE~\cite{oord2018infonce} provides a contrastive lower bound on MI with a favorable bias--variance tradeoff in relevant regimes~\cite{poole2019}, although its estimate is upper-bounded by $\log B$ for batch size $B$. Constrained-learning approaches provide a principled framework for balancing an objective against explicit constraints through adaptive Lagrange multipliers~\cite{cotter2019}. Building on these ideas, we use an InfoNCE-based leakage proxy to formulate a leakage constraint and adapt its associated privacy weight during training, while jointly studying key-conditioned signal-domain perturbation within ESCS. We propose a leakage-aware extension of ESCS for text wireless transmission with the following contributions:
\begin{itemize}
\item We introduce a key-conditioned signal-domain perturbation layer paired with a legitimate-receiver cancellation network (Sec.~\ref{sec:lapi}).
\item We augment the fixed ESCS adversarial objective with an InfoNCE-based leakage constraint whose privacy weight adapts through a dual update during training (Sec.~\ref{sec:mine}).
\item We evaluate the system over additive white Gaussian noise (AWGN) and Rayleigh fading with three training seeds, reporting bilingual evaluation understudy (BLEU) scores, the legitimate-receiver--eavesdropper gap, {InfoNCE-estimated leakage for InfoNCE-equipped configurations}, and inference overhead (Secs.~\ref{sec:setup}--\ref{sec:results}).
\item We analyze the tradeoff between eavesdropper suppression and legitimate-receiver distortion, showing that InfoNCE-guided training primarily drives operational privacy gains, while LAPI further reduces InfoNCE-estimated leakage in the combined model.
\end{itemize}

The distinction from conventional adversarial ESCS is that privacy control is not tied only to the eavesdropper's reconstruction loss: the transmitter is also trained against an explicit leakage proxy, with its privacy-penalty weight adapted during optimization. LAPI introduces a second key-dependent {mechanism} at the channel-symbol level, complementing ESCS semantic-feature encryption. Thus, leakage-aware training and signal-domain perturbation act at different parts of the communication pipeline and can be jointly incorporated into one model. Throughout the paper, non-bold ($a$), bold lowercase ($\mathbf{a}$), and bold uppercase ($\mathbf{A}$) denote scalars, vectors, and matrices, respectively. Elementwise multiplication and division are $\odot$ and $\oslash$, and $\|\cdot\|_F$ denotes the Frobenius norm.

\section{System and Threat Model}
\label{sec:system}

We adopt the ESCS architecture from~\cite{escs}, built on DeepSC~\cite{deepsc}. Here, $N$, $d_{\mathrm{model}}$, $d_c$, and $M$ denote the sequence length, embedding, channel, and key dimensions, respectively. The system comprises transmitter Alice, legitimate receiver Bob, and eavesdropper Eve. We evaluate operational privacy and statistical leakage rather than cryptographic secrecy. Alice processes a source {token sequence $\mathbf{s}$ of length $N=30$} through: (i)~embedding and positional encoding $\tilde{\mathbf{S}}=\mathrm{PE}(\mathrm{Embed}(\mathbf{s}))\in\mathbb{R}^{N\times d_{\mathrm{model}}}$; (ii)~a key-conditioned encryptor $\tilde{\mathbf{S}}_k=K_e(\tilde{\mathbf{S}},\mathbf{k})\in\mathbb{R}^{N\times d_{\mathrm{model}}}$ with shared key $\mathbf{k}\in\mathbb{R}^{M}$ ($M=128$); (iii)~a semantic encoder $\mathbf{X}_k=E(\tilde{\mathbf{S}}_k)\in\mathbb{R}^{N\times d_{\mathrm{model}}}$; (iv)~a channel encoder $\mathbf{T}_x=\mathrm{ChEnc}(\mathbf{X}_k)\in\mathbb{R}^{N\times d_c}$ with $d_c=16$; and (v)~{global RMS normalization to unit empirical average transmit power}. The transmitted signal passes through one of two wireless channels: \textit{AWGN}, $\mathbf{R}_x=\bar{\mathbf{T}}_x+\mathbf{N}$, where $\mathbf{R}_x$ is the received signal and $\mathbf{N}\in\mathbb{R}^{N\times d_c}$ is noise with {$n_{ij}\overset{\mathrm{i.i.d.}}{\sim}\mathcal{N}(0,\sigma^2)$}; or \textit{Rayleigh fading}, $\mathbf{R}_x=\mathbf{H}\odot\bar{\mathbf{T}}_x+\mathbf{N}$, where $\odot$ denotes elementwise multiplication and {$\mathbf{H}\in\mathbb{R}_+^{N\times d_c}$ contains independent Rayleigh amplitudes with scale $1/\sqrt{2}$ and $\mathbb{E}[h_{ij}^2]=1$. Each coefficient is lower-clipped at $10^{-3}$ before application and equalization.}

Bob and Eve assume perfect channel state information and apply zero-forcing equalization, {$\mathbf{R}_x\oslash\mathbf{H}=\bar{\mathbf{T}}_x+\mathbf{N}\oslash\mathbf{H}$}, where $\oslash$ denotes elementwise division; clipping limits amplification under deep fades. {During evaluation, Bob and Eve use separate forward calls with independent channel/noise realizations, so their performance gap reflects both learned behavior and independent realizations under the same channel and SNR.} Bob possesses $\mathbf{k}$ and processes the received signal through a channel decoder, decryptor $K_d(\cdot,\mathbf{k})$, and semantic decoder. Eve {uses the same channel decoder as Bob, followed by a separately parameterized attacker decoder $A(\cdot)$,} but lacks $\mathbf{k}$.

\subsection{Threat model}
\label{sec:threat}

We evaluate security under the following assumptions. {Eve uses the same channel model and SNR as Bob but, due to separate evaluation forward calls, experiences independent channel/noise realizations.} Eve has {a separately parameterized two-layer Transformer attacker decoder trained adversarially} but lacks the shared key~$\mathbf{k}$. {Thus, we consider a keyless, non-adaptive Eve with a fixed attacker architecture and do not claim robustness against an attacker adaptively retrained against the final defense or designed to suppress the signal-domain perturbation.} Stronger perturbation-aware, higher-capacity, or adaptively retrained attackers may achieve better reconstruction and are discussed in Sec.~\ref{sec:security}.

Baseline ESCS uses four round-robin training steps: $t\!\bmod\!4=0$ updates the encoder/decoder on the unencrypted path ($\mathcal{L}_{ED}$); $t\!\bmod\!4=1$ the decryptor ($\mathcal{L}_{Kd}$); $t\!\bmod\!4=2$ the attacker ($\mathcal{L}_A$); and $t\!\bmod\!4=3$ the encryptor with $\mathcal{L}_{K_e}=\mathcal{L}_{Kd}-\lambda_{\rm adv}\mathcal{L}_A$. All are cross-entropy losses, with fixed $\lambda_{\rm adv}$.

\section{Proposed Method}

\subsection{Key-conditioned signal perturbation (LAPI)}
\label{sec:lapi}

The LAPI module comprises two networks conditioned on the shared key $\mathbf{k}$. First, the \emph{perturbation generator} $P(\cdot,\mathbf{k})$ takes the normalized signal $\bar{\mathbf{T}}_x\in\mathbb{R}^{N\times d_c}$ and key $\mathbf{k}$, producing a unit-norm perturbation $\boldsymbol{\delta}\in\mathbb{R}^{N\times d_c}$:
\begin{equation}
\bar{\mathbf{T}}'_x=\bar{\mathbf{T}}_x+\rho\boldsymbol{\delta},
\quad
\boldsymbol{\delta}=P(\bar{\mathbf{T}}_x,\mathbf{k}),\;\;
\|\boldsymbol{\delta}\|_F=1,
\label{eq:lapi_gen}
\end{equation}
where $\rho>0$ controls perturbation strength. The generator is a two-layer MLP that concatenates the flattened signal ($Nd_c$ dimensions) with the key ($M$ dimensions) and {$\ell_2$-normalizes its output per sample over the $Nd_c$ dimensions}. {The perturbed signal $\bar{\mathbf{T}}'_x$ is transmitted without renormalization, so its transmit energy may differ from that of $\bar{\mathbf{T}}_x$; thus, performance changes cannot be attributed solely to the perturbation independent of energy changes.} Second, the \emph{cancellation network} $C(\cdot,\mathbf{k})$ estimates and removes the {added perturbation} at Bob before channel decoding:
\begin{equation}
\hat{\mathbf{X}}_{\mathrm{clean}}
=\mathbf{R}_x-\hat{\boldsymbol{\Delta}},
\quad
\hat{\boldsymbol{\Delta}}=C(\mathbf{R}_x,\mathbf{k}),
\label{eq:lapi_cancel}
\end{equation}
where {$\hat{\boldsymbol{\Delta}}$ estimates the added perturbation $\rho\boldsymbol{\delta}$. Under Rayleigh fading, cancellation follows perfect-channel-state-information zero-forcing equalization, so $\mathbf{R}_x$ denotes the equalized signal presented to the cancellation network.} The cleaned signal is then channel-decoded. Eve {lacks the key and is decoded without key-conditioned cancellation; a perturbation-aware or adaptively trained Eve might learn to suppress the perturbation.}

LAPI is trained with two goals: Bob should cancel the perturbation, while the perturbation should {increase Eve's reconstruction loss}. The objective is
\begin{equation}
\mathcal{L}_{\mathrm{LAPI}}=
\mathrm{MSE}(\hat{\mathbf{X}}_{\mathrm{clean}},\mathbf{X}_{\mathrm{ref}})
-\gamma\mathcal{L}_A^{(\mathrm{LAPI})},
\label{eq:lapi_loss}
\end{equation}
{where $\mathrm{MSE}$ denotes mean-squared error and $\mathbf{X}_{\mathrm{ref}}$ is an independently noise-corrupted unperturbed reference. MSE compares the cancellation output with a separately sampled noisy reference rather than isolating the perturbation under the same channel-noise realization.} The first term encourages Bob's reconstruction fidelity; the second uses Eve's cross-entropy loss on the perturbed path to degrade its reconstruction. {The perturbation generator and cancellation network are jointly optimized under this objective, while Eve is separately optimized to minimize its reconstruction cross-entropy.} The scalar $\gamma>0$ balances the two terms and is {fixed at $\gamma=0.1$}.

\subsection{InfoNCE-guided adaptive training}
\label{sec:mine}

\subsubsection{Choice of estimator and variables}
We {augment the indirect cross-entropy objective with} an InfoNCE-based leakage estimator, used as a statistical-leakage proxy for relative comparisons rather than an exact MI value. {Because each InfoNCE-equipped configuration uses separately trained projection networks, cross-configuration comparisons should be interpreted cautiously.} The leakage proxy is {$\mathcal{I}(\mathbf{X};\mathbf{R}_E)$, where $\mathbf{X}=E(\tilde{\mathbf{S}})\in\mathbb{R}^{N\times d_{\mathrm{model}}}$ is the unencrypted semantic-encoder representation and $\mathbf{R}_E\in\mathbb{R}^{N\times d_{\mathrm{model}}}$ is Eve's channel-decoded representation.} {It measures statistical dependence between these representations, not semantic recoverability, which is evaluated separately using Eve BLEU.} MINE uses the Donsker--Varadhan bound~\cite{mine}, whose variance can grow substantially with MI~\cite{smile}. We instead use InfoNCE~\cite{oord2018infonce}, which {offers a favorable bias--variance tradeoff in relevant regimes~\cite{poole2019}} and is upper-bounded by $\log B$:
\begin{equation}
\hat{\mathcal{I}}_{\mathrm{NCE}}(\mathbf{X};\mathbf{R}_E)
=\log B-\mathcal{L}_{\mathrm{NCE}},
\label{eq:infonce}
\end{equation}
where $\mathcal{L}_{\mathrm{NCE}}$ is the contrastive cross-entropy for identifying the correct pair $(\mathbf{X}_i,\mathbf{R}_{E,i})$ among $B$ candidates using
$f_\theta(\mathbf{X},\mathbf{R}_E)=
\phi_X(\mathbf{X})^\top\phi_R(\mathbf{R}_E)/\tau$,
where $\phi_X$ and $\phi_R$ are projection networks and $\tau$ is the temperature. {Each projector concatenates sequence-wise mean, maximum, and standard-deviation pooling, then applies a projection network whose output is $\ell_2$-normalized before the temperature-scaled dot product. The projectors minimize $\mathcal{L}_{\mathrm{NCE}}$ through repeated updates on the same detached minibatch, while the encryptor is trained to reduce the resulting leakage proxy.} With $B=512$, the estimate is upper-bounded by $\log512\approx6.24$ nats, where {$\log$ denotes the natural logarithm}. This ceiling is considered when interpreting absolute values in Sec.~\ref{sec:mi_analysis}.

\subsubsection{{Adaptive dual update for $\lambda$}}

We {motivate the adaptive leakage penalty through the constrained formulation}
\begin{equation}
\min_{\theta_{K_e}}\,\mathcal{L}_{\mathrm{base}}(\theta_{K_e})
\quad \text{s.t.} \quad
\hat{\mathcal{I}}_{\mathrm{NCE}}(\mathbf{X};\mathbf{R}_E)\leq\varepsilon,
\label{eq:constrained}
\end{equation}
{where $\mathcal{L}_{\mathrm{base}}$ is the baseline ESCS encryptor objective. The constraint applies to the InfoNCE leakage proxy and does not bound the true MI.} The corresponding Lagrangian is
\begin{equation}
\mathcal{L}=
\mathcal{L}_{\mathrm{base}}
+\lambda\left(\hat{\mathcal{I}}_{\mathrm{NCE}}-\varepsilon\right),
\qquad \lambda\geq0.
\end{equation}
The primal and dual updates are
\begin{align}
\theta_{K_e} &\leftarrow \theta_{K_e}-\eta_\theta
\nabla_{\theta_{K_e}}\!\left[\mathcal{L}_{\mathrm{base}}
+\lambda\hat{\mathcal{I}}_{\mathrm{NCE}}\right], \label{eq:primal}\\
\lambda &\leftarrow \mathrm{clip}\!\left(\lambda+\eta_\lambda
(\hat{\mathcal{I}}_{\mathrm{NCE}}-\varepsilon),\,
[\lambda_{\min},\lambda_{\max}]\right), \label{eq:dual}
\end{align}
where $\eta_\theta$ and $\eta_\lambda$ are the primal and dual step sizes, respectively, and $\mathrm{clip}(\cdot,[a,b])$ restricts its argument to $[a,b]$. Thus, $\lambda$ increases when $\hat{\mathcal{I}}_{\mathrm{NCE}}>\varepsilon$ and decreases when $\hat{\mathcal{I}}_{\mathrm{NCE}}<\varepsilon$.

The threshold $\varepsilon$ is {a target for the batch-dependent InfoNCE leakage proxy in nats, not a guaranteed bound on true MI}. We use $\varepsilon=0.5$ nats; {systematic sensitivity and model-selection studies for $\varepsilon$ are left for future work.}

\subsection{Implementation refinements}
\label{sec:implementation}

\subsubsection{Gated encryptor before the semantic encoder}
Algorithm~1 of~\cite{escs} applies encryption before the semantic encoder: $\mathbf{X}_k=E(K_e(\tilde{\mathbf{S}}))$. This placement {allows attention to operate directly on key-conditioned representations before the channel encoder reduces the dimension from $d_{\mathrm{model}}=128$ to $d_c=16$.} The encryptor uses a gated residual architecture:
\begin{equation}
K_e(\tilde{\mathbf{S}},\mathbf{k})
=(1-\mathbf{G})\odot\tilde{\mathbf{S}}
+\mathbf{G}\odot\mathrm{MLP}([\tilde{\mathbf{S}},\mathbf{k}]),
\label{eq:gated}
\end{equation}
where $\mathbf{G}=\sigma(\mathbf{W}_g[\tilde{\mathbf{S}},\mathbf{k}]+\mathbf{b}_g)\in\mathbb{R}^{N\times d_{\mathrm{model}}}$ is a per-dimension gate, $\mathbf{W}_g\in\mathbb{R}^{d_{\mathrm{model}}\times(d_{\mathrm{model}}+M)}$, $\mathbf{b}_g\in\mathbb{R}^{d_{\mathrm{model}}}$, and the MLP is a three-layer network with Gaussian error linear unit (GELU) and LayerNorm. {The key $\mathbf{k}\in\mathbb{R}^{M}$ is broadcast across the $N$ token positions before concatenation. $\mathbf{W}_g$ and $\mathbf{b}_g$ are zero-initialized, so every entry of $\mathbf{G}$ is initially $0.5$, equally weighting the residual and MLP branches.}

\subsubsection{Full encryptor loss}
The full encryptor loss combines the baseline adversarial objective with an InfoNCE penalty when active and cancellation reconstruction when LAPI is active:
\begin{equation}
\mathcal{L}_{K_e}^{\mathrm{full}}
=\mathcal{L}_{Kd}
-\lambda_{\mathrm{adv}}\mathcal{L}_A
+\frac{\lambda}{\log B}\max(0,\hat{\mathcal{I}}_{\mathrm{NCE}})
{+\beta_{\mathrm{rec}}\mathcal{L}_{\mathrm{MSE}}},
\label{eq:full_loss}
\end{equation}
{where the final term is included only with LAPI, with $\beta_{\mathrm{rec}}=0.05$.} Here, $\lambda_{\mathrm{adv}}=0.5$ is the fixed adversarial weight and $\lambda$ is the adaptive multiplier from~\eqref{eq:dual}. {The implementation uses $\max(0,\hat{\mathcal{I}}_{\mathrm{NCE}})$ in the encryptor penalty. The threshold $\varepsilon$ is absent from the primal gradient because $-\lambda\varepsilon$ is constant with respect to the encryptor parameters and instead enters the dual update.} {Division by $\log B$ normalizes the penalty by the finite-batch InfoNCE ceiling.}

\begin{figure*}[t]
\centering
\includegraphics[width=1.1\columnwidth]{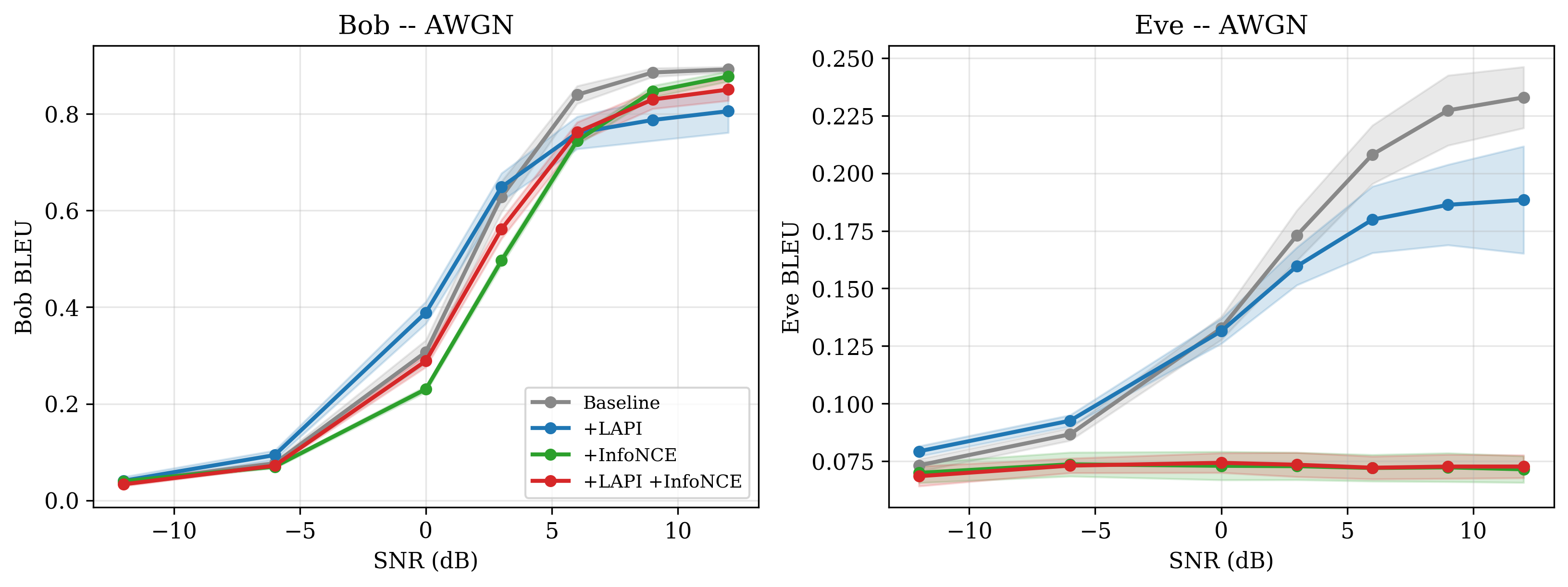}
\caption{Composite BLEU versus AWGN SNR for all configurations. Lines show three-seed means; bands show $\pm1$ standard deviation. InfoNCE-equipped models {maintain low Eve BLEU at high SNR}.}
\label{fig:bleu_awgn}
\end{figure*}

\section{Experimental Setup}
\label{sec:setup}

\subsection{Dataset, preprocessing, and model configuration}
We use the Europarl English corpus following~\cite{escs,deepsc}. {Tokenization, vocabulary construction, padding, and masking follow the inherited ESCS/DeepSC preprocessing pipeline.} Sentences are tokenized and truncated to $N=30$ tokens. The training and test sets contain $1{,}473{,}206$ and $2{,}000$ sentences, respectively. {The vocabulary contains $64{,}742$ tokens.}

The {semantic encoder and decoder} use a Transformer with $d_{\mathrm{model}}=128$, $N_{\mathrm{layers}}=4$, $N_{\mathrm{heads}}=8$, and $d_{\mathrm{ff}}=512$; $d_c=16$ and $M=128$. {Eve's attacker decoder uses a two-layer Transformer.} The gated encryptor and decryptor are three-layer MLPs ($2d_{\mathrm{model}}\to4d_{\mathrm{model}}\to4d_{\mathrm{model}}\to d_{\mathrm{model}}$) with Gaussian error linear unit (GELU) activations and LayerNorm. {Since $M=d_{\mathrm{model}}=128$, the input dimension is $2d_{\mathrm{model}}=256$, with the key broadcast across $N$ positions before concatenation.} LAPI uses two-layer MLPs with dimensions {$608\to480\to480$, where $608=Nd_c+M$ and the $480=Nd_c$ output is reshaped to $N\times d_c$}, with $\gamma=0.1$ in~\eqref{eq:lapi_loss} and {$\rho=\min(\exp(\ell_\rho),0.5)$, initialized at $0.3$}. InfoNCE projectors $\phi_X,\phi_R$ are two-layer MLPs ($3d_{\mathrm{model}}\to512\to256$) with sequence-wise mean, max, and standard-deviation pooling and LayerNorm. {Their outputs are $\ell_2$-normalized before the temperature-scaled dot product.}

\subsection{Training, evaluation protocol, and metrics}

We use {PyTorch Adam} ($\beta_1\!=\!0.9$, $\beta_2\!=\!0.98$, $\epsilon_{\mathrm{Adam}}\!=\!10^{-8}$, \texttt{weight\_decay}$=5\!\times\!10^{-4}$) with learning rate $10^{-4}$ and batch size $B=512$. Each configuration is trained for $6{,}000$ rounds at $\mathrm{SNR}\!=\!10$\,dB. {The baseline follows the four-step ESCS round-robin updates in Sec.~\ref{sec:threat}; configurations with LAPI and/or InfoNCE additionally perform their corresponding steps when active, rather than using a common six-step cycle. During training, each batch uses either AWGN or Rayleigh fading with equal probability, yielding one checkpoint that is evaluated separately on both channels.} The fixed adversarial weight is $\lambda_{\mathrm{adv}}=0.5$. For InfoNCE-guided training, {$\lambda$ is initialized at $0.2$ and held fixed for the first $500$ rounds; dual updates begin thereafter.} $\lambda$ is then clipped to $[0.05,1.5]$, with leakage target $\varepsilon=0.5$\,nats, dual step $\eta_\lambda=10^{-3}$, InfoNCE temperature $\tau=0.1$, and $K\!=\!10$ critic updates per round. {Each critic phase updates only the InfoNCE projectors $\phi_X$ and $\phi_R$ for $10$ steps on the same detached minibatch representation, with communication-model parameters fixed.}

{All configurations initialize the common semantic backbone from the same pretrained checkpoint. The associated encoder/decoder token embeddings are then frozen during training.} Every configuration is {trained from the common pretrained initialization} with three random seeds ($0$, $1$, $2$). All BLEU, security-gap, and MI quantities are reported as mean $\pm$ standard deviation across seeds. Evaluation uses AWGN and Rayleigh channels at SNR $\in\{-12,-6,0,3,6,9,12\}$\,dB. {The reported SNR sets the channel-noise level relative to the normalized signal before LAPI perturbation; since the perturbed signal is not renormalized, it does not control for post-perturbation transmit energy.} Training/evaluation use a single NVIDIA A100-40GB GPU with bfloat16 mixed precision and the PyTorch \texttt{scaled\_dot\_product\_attention} kernel.

We report (i) the composite BLEU score~\cite{bleu} ($0.6\!\cdot\!\text{BLEU-1}+0.4\!\cdot\!\text{BLEU-2}$) for Bob and Eve, {following our ESCS implementation}; (ii) the {operational security gap, defined as the empirical BLEU difference $\mathrm{Bob}-\mathrm{Eve}$, not an information-theoretic secrecy measure}; (iii) the privacy-efficiency ratio
\begin{equation}
\mathrm{PE}=\frac{\Delta\mathrm{Eve}}{|\Delta\mathrm{Bob}|+\epsilon_0},
\label{eq:pe}
\end{equation}
where $\epsilon_0=10^{-3}$, {$\Delta\mathrm{Eve}=\mathrm{Eve}_{\mathrm{baseline}}-\mathrm{Eve}_{\mathrm{method}}$ and $\Delta\mathrm{Bob}=\mathrm{Bob}_{\mathrm{method}}-\mathrm{Bob}_{\mathrm{baseline}}$ at the same SNR. Positive $\Delta\mathrm{Eve}$ denotes reduced Eve BLEU. PE is descriptive and sensitive to $\epsilon_0$ when $|\Delta\mathrm{Bob}|$ approaches zero}; and (iv) the InfoNCE leakage proxy $\hat{\mathcal{I}}_{\mathrm{NCE}}(\mathbf{X};\mathbf{R}_E)$ in nats. It is reported only for InfoNCE-equipped configurations (\emph{+InfoNCE} and \emph{Combined}), which contain trained projectors $\phi_X,\phi_R$. {Because projectors are trained separately per configuration, cross-configuration leakage differences should be interpreted cautiously.} We omit baseline leakage because baseline ESCS does not train InfoNCE projectors; a common estimator is left for future work.

\section{Results and Discussion}
\label{sec:results}

\subsection{BLEU performance and security gap}

Fig.~\ref{fig:bleu_awgn} shows Bob and Eve composite BLEU versus AWGN SNR, with $\pm$one-standard-deviation bands over three seeds. For InfoNCE-equipped models, Eve BLEU remains {low at positive SNR, reaching about $0.07$ at high SNR, indicating persistent eavesdropper suppression as noise decreases.} Tables~\ref{tab:awgn} and~\ref{tab:rayleigh} report results at $6$, $9$, and $12$\,dB for AWGN and Rayleigh. At $12$\,dB AWGN, \emph{+InfoNCE} gives the largest operational gap ($0.806$), while \emph{Combined} gives the lowest {InfoNCE leakage proxy} ($0.747$ vs.\ $0.817$\,nats). {Since projectors are trained separately, leakage values are proxy comparisons rather than common-estimator estimates.} This indicates a small tradeoff between {operational gap and estimated leakage}. At $12$\,dB Rayleigh, \emph{Combined} gives a {gap of $0.600$ versus $0.587$ for \emph{+InfoNCE}, interpreted descriptively over three seeds. This is consistent with greater LAPI utility under residual channel variability after equalization.} {Seed variability is small, with $\sigma_{\mathrm{Eve}}\le0.013$ for reported high-SNR Eve BLEU.}

LAPI alone is not uniformly better: at $12$\,dB AWGN, \emph{+LAPI} reduces Eve BLEU by $0.045$ but Bob BLEU by $0.087$, narrowing the security gap from $0.659$ to $0.617$. In contrast, \emph{+InfoNCE} reduces Eve BLEU by $0.162$ with only a $0.015$ Bob reduction, yielding $\mathrm{PE}\approx0.51$ for \emph{+LAPI} versus {$10.13$} for \emph{+InfoNCE}; {the same PE ordering holds at $6$ and $9$\,dB.} The Bob degradation {is consistent with residual distortion from imperfect perturbation cancellation, though other reconstruction errors are not isolated.} LAPI is most valuable with InfoNCE-guided training: \emph{Combined} retains high Bob BLEU ($0.850$ at $12$\,dB) while {showing lower InfoNCE leakage than \emph{+InfoNCE} ($0.747$ vs.\ $0.817$\,nats), although separately trained projectors make this a proxy comparison.} Thus, \emph{Combined} provides a more privacy-conservative point, trading a small operational gap for lower {InfoNCE-estimated leakage}.

\begin{table}[t]
\centering
\caption{AWGN: composite BLEU, {operational BLEU gap}, and InfoNCE {leakage proxy} (nats) at $6$, $9$, $12$\,dB. Mean $\pm$ standard deviation over three seeds. Best Eve, gap, and {InfoNCE} per SNR in \textbf{bold}; lower Eve BLEU indicates stronger {eavesdropper suppression}. {InfoNCE is reported only for equipped configurations using separately trained projectors.}}
\label{tab:awgn}
\scriptsize
\setlength{\tabcolsep}{2.5pt}
\begin{tabular}{lcccc}
\toprule
Config & Bob & Eve & Gap & {InfoNCE}\\
\midrule
\multicolumn{5}{c}{\textbf{6\,dB}}\\
Baseline       & 0.839\,$\pm$\,0.018 & 0.208\,$\pm$\,0.013 & 0.631\,$\pm$\,0.013 & n/a\\
+LAPI          & 0.761\,$\pm$\,0.033 & 0.180\,$\pm$\,0.014 & 0.581\,$\pm$\,0.046 & n/a\\
+InfoNCE       & 0.744\,$\pm$\,0.009 & \textbf{0.072\,$\pm$\,0.006} & 0.672\,$\pm$\,0.014 & 0.748\,$\pm$\,0.067\\
Combined       & 0.762\,$\pm$\,0.021 & \textbf{0.072\,$\pm$\,0.005} & \textbf{0.690\,$\pm$\,0.017} & \textbf{0.707\,$\pm$\,0.055}\\
\midrule
\multicolumn{5}{c}{\textbf{9\,dB}}\\
Baseline       & 0.886\,$\pm$\,0.009 & 0.227\,$\pm$\,0.015 & 0.658\,$\pm$\,0.012 & n/a\\
+LAPI          & 0.787\,$\pm$\,0.043 & 0.186\,$\pm$\,0.017 & 0.601\,$\pm$\,0.060 & n/a\\
+InfoNCE       & 0.846\,$\pm$\,0.012 & 0.072\,$\pm$\,0.006 & \textbf{0.774\,$\pm$\,0.018} & 0.828\,$\pm$\,0.044\\
Combined       & 0.830\,$\pm$\,0.019 & \textbf{0.073\,$\pm$\,0.005} & 0.757\,$\pm$\,0.015 & \textbf{0.720\,$\pm$\,0.054}\\
\midrule
\multicolumn{5}{c}{\textbf{12\,dB}}\\
Baseline       & 0.892\,$\pm$\,0.005 & 0.233\,$\pm$\,0.013 & 0.659\,$\pm$\,0.010 & n/a\\
+LAPI          & 0.805\,$\pm$\,0.044 & 0.188\,$\pm$\,0.023 & 0.617\,$\pm$\,0.068 & n/a\\
+InfoNCE       & 0.877\,$\pm$\,0.010 & \textbf{0.071\,$\pm$\,0.006} & \textbf{0.806\,$\pm$\,0.016} & 0.817\,$\pm$\,0.032\\
Combined       & 0.850\,$\pm$\,0.023 & 0.073\,$\pm$\,0.005 & 0.777\,$\pm$\,0.020 & \textbf{0.747\,$\pm$\,0.082}\\
\bottomrule
\end{tabular}
\end{table}

\begin{table}[t]
\centering
\caption{Rayleigh fading: composite BLEU, {operational BLEU gap}, and InfoNCE {leakage proxy} (nats) at $6$, $9$, $12$\,dB. {Mean $\pm$ standard deviation over three seeds. InfoNCE is reported only for equipped configurations using separately trained projectors.}}
\label{tab:rayleigh}
\scriptsize
\setlength{\tabcolsep}{2.5pt}
\begin{tabular}{lcccc}
\toprule
Config & Bob & Eve & Gap & {InfoNCE}\\
\midrule
\multicolumn{5}{c}{\textbf{6\,dB}}\\
Baseline       & 0.366\,$\pm$\,0.022 & 0.133\,$\pm$\,0.004 & 0.233\,$\pm$\,0.019 & n/a\\
+LAPI          & 0.398\,$\pm$\,0.018 & 0.132\,$\pm$\,0.005 & \textbf{0.266\,$\pm$\,0.020} & n/a\\
+InfoNCE       & 0.289\,$\pm$\,0.004 & \textbf{0.074\,$\pm$\,0.006} & 0.215\,$\pm$\,0.008 & 0.321\,$\pm$\,0.062\\
Combined       & 0.326\,$\pm$\,0.010 & 0.074\,$\pm$\,0.005 & 0.252\,$\pm$\,0.007 & \textbf{0.315\,$\pm$\,0.028}\\
\midrule
\multicolumn{5}{c}{\textbf{9\,dB}}\\
Baseline       & 0.580\,$\pm$\,0.023 & 0.167\,$\pm$\,0.010 & 0.413\,$\pm$\,0.019 & n/a\\
+LAPI          & 0.580\,$\pm$\,0.023 & 0.155\,$\pm$\,0.011 & 0.425\,$\pm$\,0.032 & n/a\\
+InfoNCE       & 0.481\,$\pm$\,0.014 & 0.073\,$\pm$\,0.006 & 0.407\,$\pm$\,0.016 & 0.549\,$\pm$\,0.047\\
Combined       & 0.516\,$\pm$\,0.015 & \textbf{0.074\,$\pm$\,0.004} & \textbf{0.443\,$\pm$\,0.011} & \textbf{0.507\,$\pm$\,0.044}\\
\midrule
\multicolumn{5}{c}{\textbf{12\,dB}}\\
Baseline       & 0.745\,$\pm$\,0.013 & 0.193\,$\pm$\,0.013 & 0.552\,$\pm$\,0.014 & n/a\\
+LAPI          & 0.690\,$\pm$\,0.022 & 0.173\,$\pm$\,0.014 & 0.517\,$\pm$\,0.034 & n/a\\
+InfoNCE       & 0.659\,$\pm$\,0.010 & \textbf{0.072\,$\pm$\,0.006} & 0.587\,$\pm$\,0.015 & 0.675\,$\pm$\,0.020\\
Combined       & 0.673\,$\pm$\,0.020 & 0.073\,$\pm$\,0.004 & \textbf{0.600\,$\pm$\,0.016} & \textbf{0.638\,$\pm$\,0.049}\\
\bottomrule
\end{tabular}
\end{table}

\subsection{Information-leakage analysis}
\label{sec:mi_analysis}

Adding LAPI to the InfoNCE-guided objective yields {lower InfoNCE leakage proxies at all reported high-SNR points}. On AWGN, leakage decreases from $0.817$ to $0.747$\,nats at $12$\,dB ($8.6\%$), with the same trend at $6$ and $9$\,dB {($0.748$ to $0.707$ and $0.828$ to $0.720$\,nats)}. On Rayleigh, {the trend also holds at all reported points}; at $12$\,dB, leakage decreases from $0.675$ to $0.638$\,nats ($5.5\%$). {These differences use separately trained projectors and should be interpreted as proxy comparisons.} The InfoNCE ceiling is $\log B\approx6.24$ nats for $B=512$, well above the reported $\sim0.3$--$0.8$\,nats, so {ceiling saturation is not apparent}. Since projectors exist only in InfoNCE-equipped configurations, baseline comparisons rely on the BLEU gaps in Tables~\ref{tab:awgn} and~\ref{tab:rayleigh}.

\subsection{Ablation at 6\,dB}

\begin{figure*}[t]
\centering
\includegraphics[width=1.1\columnwidth]{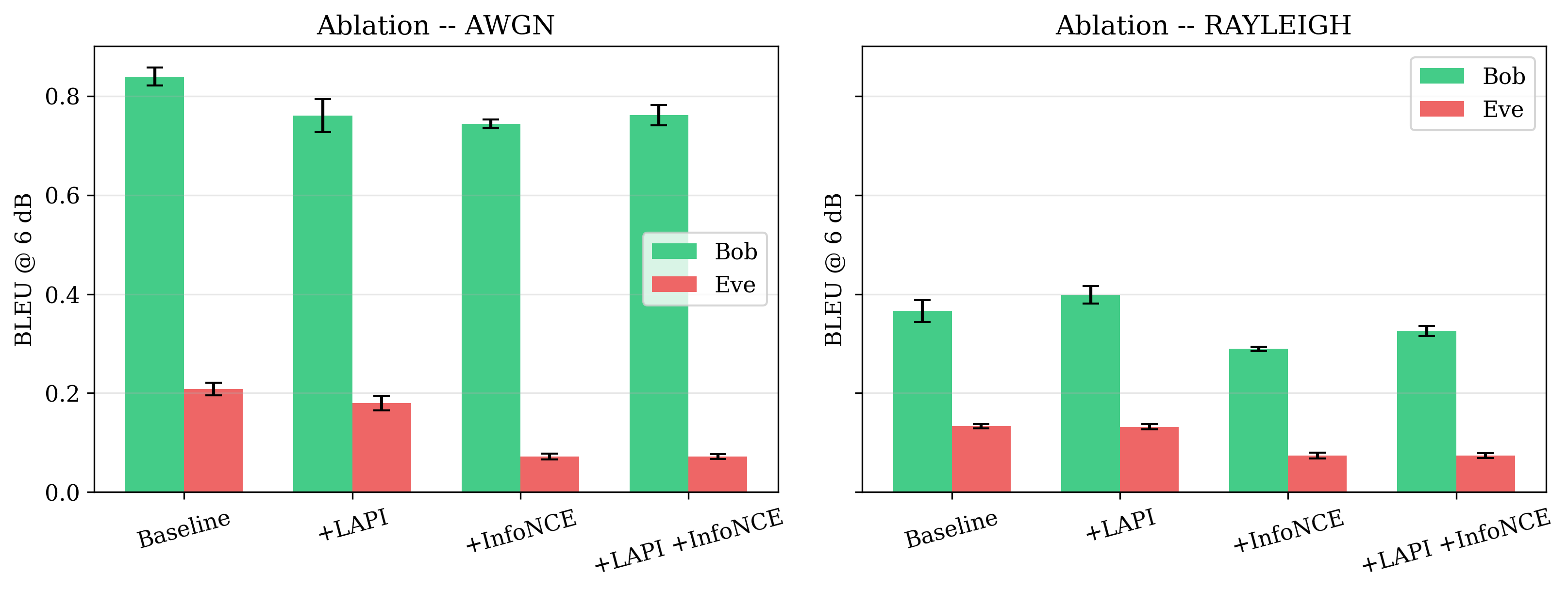}
\caption{Ablation at $6$\,dB on AWGN (left) and Rayleigh (right). Error bars: $\pm 1$ standard deviation over three seeds.}
\label{fig:ablation}
\end{figure*}

Fig.~\ref{fig:ablation} compares Bob and Eve BLEU at $6$\,dB across all four configurations and both channels. Eve BLEU is {identical for the two InfoNCE-equipped configurations: $0.072$ on AWGN and $0.074$ on Rayleigh}, while \emph{Combined} improves Bob BLEU over \emph{+InfoNCE} from {$0.744$ to $0.762$ on AWGN and $0.289$ to $0.326$ on Rayleigh}. {LAPI and InfoNCE operate at different parts of the pipeline: LAPI perturbs the transmitted signal, whereas InfoNCE adds a training-time leakage penalty.} At $6$\,dB, {combining them preserves the low Eve BLEU of \emph{+InfoNCE} while improving Bob BLEU on both channel models.}

\subsection{Complexity and inference overhead}
\label{sec:complexity}

Table~\ref{tab:complexity} reports parameter counts and per-batch ($B=512$) forward latency on an NVIDIA A100-40GB. LAPI adds $1.05$\,M parameters ($+2.3\%$) and $\approx0.05$\,ms {($+1.3\%$)} AWGN latency. {The InfoNCE projectors and adaptive-$\lambda$ update are training-only and add no inference-time components or computation.} The combined configuration has {latency close to \emph{+LAPI} ($4.04$ vs.\ $4.02$\,ms on AWGN; $4.18$ vs.\ $4.22$\,ms on Rayleigh). Rayleigh latency is up to $0.18$\,ms higher than AWGN.} Overall, {LAPI adds little measured inference overhead, while InfoNCE adds none.}

\begin{table}[t]
\centering
\caption{Parameter counts and forward latency per $512$-sentence batch on an NVIDIA A100-40GB. {InfoNCE projectors are used only in training but included in the reported parameter counts.}}
\label{tab:complexity}
\scriptsize
\setlength{\tabcolsep}{3pt}
\begin{tabular}{lcccccc}
\toprule
 & Total & Crypto & LAPI & InfoNCE & \multicolumn{2}{c}{Fwd latency (ms)}\\
\cmidrule(lr){6-7}
Config & (M) & (K) & (K) & (K) & AWGN & Rayleigh\\
\midrule
Baseline       & 45.11 & 989.7 & n/a    & n/a    & 3.97 & 4.13\\
+LAPI          & 46.15 & 989.7 & 1046.4 & n/a    & 4.02 & 4.22\\
+InfoNCE       & 46.16 & 989.7 & n/a    & 1052.7 & 3.58 & 3.71\\
Combined       & 47.21 & 989.7 & 1046.4 & 1052.7 & 4.04 & 4.18\\
\bottomrule
\end{tabular}
\end{table}

\subsection{Limitations and threat-model discussion}
\label{sec:security}

We discuss the proposed enhancements under the threat model of Sec.~\ref{sec:threat}. InfoNCE is used only during training and {adds no inference-time component or computation}, whereas LAPI deploys the key-conditioned perturbation generator at Alice and cancellation network at Bob.

\emph{LAPI: protection and limits.} {Under the evaluated non-adaptive, keyless-Eve model,} $\boldsymbol{\delta}=P(\bar{\mathbf{T}}_x,\mathbf{k})$ acts as structured interference that Eve cannot reproduce from the key and degrades its reconstruction, especially with InfoNCE-guided training. We do not claim cryptographic indistinguishability: an adaptive Eve may learn to suppress the perturbation, and perturbation-aware attacks are not evaluated.

\emph{InfoNCE-guided training and estimator limits.} {The objective trains the encryptor to reduce InfoNCE-estimated leakage between the unencrypted semantic representation $\mathbf{X}=E(\tilde{\mathbf{S}})$ and Eve's channel-decoded representation $\mathbf{R}_E$, with $\lambda$ adapting the leakage penalty.} As a learned MI lower bound available only for InfoNCE-equipped configurations, it neither permits direct baseline MI comparison nor provides Shannon or cryptographic secrecy guarantees. A common estimator and sensitivity analysis over $\varepsilon$ are left for future work.

\emph{Combined effect and attacker assumptions.} InfoNCE-guided training is the primary source of operational eavesdropper suppression, while LAPI {is associated with lower InfoNCE leakage in the combined model, subject to separately trained estimators, and can distort Bob's reconstruction when used alone.} \emph{Combined} has {the lowest leakage proxy at all reported high-SNR points}. At {$9$ and $12$\,dB AWGN, \emph{+InfoNCE} has the larger Bob--Eve gap, while \emph{Combined} does so at $6$\,dB and all reported Rayleigh points.} Eve is a {separately parameterized, non-adaptive two-layer Transformer}; higher-capacity, perturbation-aware, or retrained attackers are not evaluated.

Extensions to multiple-input multiple-output (MIMO) and more complex propagation~\cite{ismail2026thz}, multi-user wiretap settings, and image/video modalities are left for future work.

\section{Conclusion}

This work advances privacy-aware semantic communication by combining protection at the signal level with leakage-aware training within a unified encrypted framework. We introduced two enhancements to ESCS~\cite{escs}: LAPI adds a {key-conditioned signal-domain perturbation at Alice with cancellation at Bob}, while InfoNCE-guided training {augments the adversarial objective with an adaptive contrastive leakage penalty}. Across AWGN and Rayleigh fading and three training seeds, InfoNCE-guided training is the primary driver of eavesdropper suppression, reducing Eve BLEU at $12$\,dB from $0.233\pm0.013$ to $0.071\pm0.006$ on AWGN and from $0.193\pm0.013$ to $0.072\pm0.006$ on Rayleigh. Adding LAPI {is associated with $5$--$9\%$ lower InfoNCE leakage at the reported high-SNR points, although separately trained projectors warrant cautious comparison. LAPI adds about $1.3\%$ measured AWGN forward latency over baseline but can cause substantial Bob distortion when used alone. Mixed-channel training yields one checkpoint evaluated separately on AWGN and Rayleigh.} Future work includes adaptive and perturbation-aware attackers, a common InfoNCE estimator, MIMO and multi-user wiretap channels, and tighter MI estimators.



\begin{thebibliography}{00}

\bibitem{deepsc}
H.~Xie, Z.~Qin, G.~Y.~Li, and B.-H.~Juang, ``Deep learning enabled semantic communication systems,'' \emph{IEEE Trans. Signal Process.}, vol.~69, pp.~2663--2675, 2021.

\bibitem{qin2022}
Z.~Qin, X.~Tao, J.~Lu, W.~Tong, and G.~Y.~Li,
``Semantic communications: Principles and challenges,''
\emph{arXiv:2201.01389}, 2022.

\bibitem{escs}
X.~Luo, Z.~Chen, M.~Tao, and F.~Yang, ``Encrypted semantic communication using adversarial training for privacy preserving,'' \emph{IEEE Commun. Lett.}, vol.~27, no.~6, pp.~1486--1490, Jun. 2023.

\bibitem{gan}
I.~Goodfellow \emph{et al.}, ``Generative adversarial nets,'' in \emph{Proc. Adv. Neural Inf. Process. Syst. (NeurIPS)}, vol.~27, 2014.

\bibitem{abadi2016}
M.~Abadi and D.~G.~Andersen, ``Learning to protect communications with adversarial neural cryptography,'' \emph{arXiv:1610.06918}, 2016.

\bibitem{sagduyu2025}
Y.~E.~Sagduyu, T.~Erpek, A.~Yener, and S.~Ulukus,
``Privacy-preserving semantic communications via multi-task learning and adversarial perturbations,''
\emph{IEEE Network}, 2026.

\bibitem{shao2025}
W.~Chen, Q.~Yang, S.~Shao, S.~Tang, Z.~Shi, and S.~Yu,
``Privacy-preserving semantic communication over wiretap channels with learnable differential privacy,''
\emph{arXiv:2510.23274}, 2025.

\bibitem{he2024}
X.~Ren, J.~Wu, H.~Xu, and X.~Chen,
``Diffusion model based secure semantic communications with adversarial purification,''
in \emph{Proc. IEEE 10th Conf. Big Data Security on Cloud},
pp.~130--134, 2024.

\bibitem{tang2025}
S.~Tang \emph{et~al.}, ``Towards secure semantic communications in the presence of intelligent eavesdroppers,'' \emph{arXiv:2503.23103}, 2025.

\bibitem{mine}
M.~I.~Belghazi \emph{et al.}, ``Mutual information neural estimation,'' in \emph{Proc. Int. Conf. Mach. Learn. (ICML)}, vol.~80, pp.~531--540, 2018.

\bibitem{alhaj2026signdeepsc}
K.~Alhaj, R.~Tajeddine, and H.~Sarieddeen,
``SignDeepSC: A semantic signature-based approach for robust semantic communication,''
\emph{arXiv preprint arXiv:2607.25676}, 2026.

\bibitem{smile}
J.~Song and S.~Ermon, ``Understanding the limitations of variational mutual information estimators,'' in \emph{Proc. ICLR}, 2020.

\bibitem{oord2018infonce}
A.~van~den~Oord, Y.~Li, and O.~Vinyals, ``Representation learning with contrastive predictive coding,'' \emph{arXiv:1807.03748}, 2018.

\bibitem{poole2019}
B.~Poole, S.~Ozair, A.~van~den~Oord, A.~A.~Alemi, and G.~Tucker,
``On variational bounds of mutual information,'' in \emph{Proc. ICML},
vol.~97, pp.~5171--5180, 2019.

\bibitem{cotter2019}
A.~Cotter, H.~Jiang, M.~R.~Gupta \emph{et~al.}, ``Optimization with non-differentiable constraints with applications to fairness, recall, churn, and other goals,'' \emph{J. Mach. Learn. Res.}, vol.~20, no.~172, pp.~1--59, 2019.

\bibitem{bleu}
K.~Papineni, S.~Roukos, T.~Ward, and W.-J.~Zhu, ``BLEU: A method for automatic evaluation of machine translation,'' in \emph{Proc. ACL}, pp.~311--318, 2002.

\bibitem{ismail2026thz}
F.~Ismail, H.~Sarieddeen, and J.~Fahs, ``Semantic communications in the THz band,'' \emph{arXiv preprint arXiv:2607.07455}, 2026.

\end{thebibliography}
\end{document}